\documentclass[%
 reprint,
 amsmath,amssymb,
 aps,
]{revtex4-1}

\usepackage{graphicx}
\usepackage{dcolumn}
\usepackage{bm}

\begin{document}

\preprint{APS/123-QED}

\title{Sequence  Heterogeneity Accelerates  Protein Search for Targets on DNA}

\author{Alexey A. Shvets}
\author{Anatoly  B.  Kolomeisky}%
\email{tolya@rice.edu}
\affiliation{Rice University, Department of Chemistry and Center for Theoretical Biological Physics, Houston, Texas 77005, USA}




\date{\today}

\begin{abstract}
The process of protein search for specific binding sites on DNA is fundamentally important since it marks the beginning of  all major biological processes. We present a theoretical investigation that probes the role of DNA sequence symmetry, heterogeneity and chemical composition in the protein search dynamics. Using a discrete-state stochastic approach with a first-passage events analysis, which takes into account the most relevant physical-chemical processes, a full analytical description of the search dynamics is obtained. It is found that, contrary to existing views, the protein search is generally faster on DNA with more heterogeneous sequences. In addition, the search dynamics might be affected by the chemical composition near the target site. The physical origins of these phenomena are discussed. Our results suggest that biological processes might be effectively regulated by modifying chemical composition, symmetry and heterogeneity of a genome.

\begin{description}
\item[Usage]
Secondary publications and information retrieval purposes.
\item[PACS numbers]
May be entered using the \verb+\pacs{#1}+ command.
\item[Structure]
You may use the \texttt{description} environment to structure your abstract;
use the optional argument of the \verb+\item+ command to give the category of each item. 
\end{description}
\end{abstract}

\pacs{Valid PACS appear here}
\maketitle



Many biological processes are initiated by proteins binding the specific target sequences on DNA. In particular, this process is responsible for transferring and maintaining the genetic information contained in DNA \cite{alberts,lodish,phillips}. It was recognized long time ago that finding these specific binding sites could be quite a complicated task due to large number of other nonspecific sites ($\simeq 10^6-10^9$) and low concentration of relevant proteins. But experiments suggest that many proteins find their targets much faster than expected from 3D bulk diffusion estimates \cite{riggs70,halford04,vandenbroek08,mirny09,kolomeisky11}. This surprising phenomenon is known as a {\it facilitated diffusion}. A significant progress in explaining facilitated diffusion processes has been achieved in recent years due to multiple experimental and theoretical advances \cite{halford04,vandenbroek08,mirny09,kolomeisky11,berg81,berg85,winter81,gowers05,iwahara06,kolesov07,wang06,elf07,tafvizi08,hu06,rau10,larson11,hammar12,zandarashvili12,veksler13,marcovitz13,koslover11,sheinman12,landry13,tafvizi11,leith12,kolomeisky12,esadze14,bauer13}. However, the detailed mechanisms of the protein search for targets on DNA  remain not well understood \cite{mirny09,kolomeisky11,veksler13}.

It is now widely accepted that proteins searching for the specific binding sites on DNA at some conditions might alternate between 3D and 1D search modes \cite{berg81,winter81,halford04,mirny09,kolomeisky11}. This means that  the protein molecule binds nonspecifically to DNA, then slides along the chain, unbinds and repeats the scanning cycle several times until it finds the target. Recent single-molecule experiments that can visualize the dynamics of individual molecules  support this picture \cite{gowers05,wang06,elf07,tafvizi11,hammar12,landry13}. These observations also underline the critical role of protein-DNA interactions in the facilitated diffusion. Since DNA molecule is a heterogeneous biopolymer, the sequence symmetry and its chemical composition must be an important factor in the protein search for targets. However, how specifically the sequence heterogeneity influences the protein search dynamics remains a controversial problem.

The protein search  on the random DNA sequences have been theoretically investigated before \cite{mirny09,shklovskii06pre}. Comparing this process with a motion in the random potential, it was shown that the heterogeneous character of the chain leads to the larger search times in comparison with a homogeneous case. But later it was argued that this result is not applicable to the protein search \cite{veksler13}. It is just an artifact of the  continuum  approximation, which assumed that the protein can reach the target only via DNA sliding, neglecting 3D associations and dissociations events \cite{veksler13}.  A more advanced computational study of the sequence heterogeneity also found that it usually slows down the facilitated diffusion by creating traps \cite{brackley12}. But it was also suggested that the properly positioned traps in the funnel shape near the target can accelerate the protein search \cite{brackley12}.  At the same time, it is not clear if such funnel distributions are observed in real systems. Furthermore, recent theoretical studies of Lukatsky and coworkers \cite{afek11,afek12,afek13,lukatsky14} suggested that the sequence symmetry creates additional effective interactions between DNA and protein molecules. Using methods of equilibrium statistical mechanics, it was found that more homogeneous segments of DNA effectively attract proteins stronger than the heterogeneous segments. However, the role of these effective interactions in the protein search for targets on DNA has not been tested yet.

In this article, we present a theoretical approach that allows us to investigate explicitly the effect of sequence heterogeneity in the protein search for targets on DNA. It is based on a discrete-state stochastic method which takes into account the most relevant physical-chemical processes of the protein search by analyzing first-passage events in the system \cite{kolomeisky12,veksler13}. The advantage of this method is that it provides a full analytical description of the facilitated diffusion. One of the main results of this approach is a development of the general dynamic phase diagram for the target search \cite{veksler13}. Three dynamic search regimes where identified depending on the different length scales in the system.  For protein sliding length $\lambda$ larger that the size of the DNA chain $L$, the protein molecule always stays on DNA and performs 1D search with a random-walk dynamics. This leads to the quadratic scaling of the search times as a function of the DNA length. When the sliding length is smaller than the length of DNA but larger than the target size ($1 < \lambda < L$), the protein is searching by combining 3D and 1D motions. In this sliding regime, the linear scaling of the search times is observed. A different dynamic phase is found for the case of the sliding length smaller than the target size, $\lambda < 1$. Here the search is accomplished only via 3D associations and dissociations events without sliding along the DNA molecule. This also leads to the linear scaling in the search times as a function of the DNA length.

In our model, we consider a single DNA molecule  with $L+1$ binding sites and a single protein molecule, as shown in Fig. 1.  One of the binding sites is a target, and for convenience we put it in the middle of the chain, i.e., $m=L/2+1$. To model the sequence heterogeneity, we assume that each monomer in the DNA chain can be in one of two chemical states, $A$ or $B$ (see Fig. 1). When the protein is bound to the segment $A$ ($B$) it interacts with energy $\varepsilon^{A}$ ($\varepsilon^{B}$), and $\varepsilon = \varepsilon^A-\varepsilon^B\ge 0$. This means that the protein attracts stronger to the $B$ sites. The protein molecule can diffuse along DNA with the rate $u_{A}\equiv u$ ($u_{B}=ue^{-\varepsilon}$, where $\varepsilon$ is measured in $k_BT$ units). Here we assume that, independently of the chemical state of their neighbors, moving out of the sites $A$ are characterized by the rate  $u_{A}$, while the diffusion out of the sites  $B$ is given by $u_{B}$. The protein search starts in the solution that we label as a state $0$. Then the protein molecule can bind to any site $A$ or $B$ on DNA with the corresponding rates $k_{on}^{A}\equiv k_{on}$ or $k_{on}^{B}=k_{on}e^{\theta\varepsilon}$. Similarly, the dissociations from the DNA chain are described by the rates $k_{off}^{A}\equiv k_{off}$  and $k_{off}^{B}=k_{off}e^{(\theta-1)\varepsilon}$. Here the parameter $0 \le \theta \le 1$ specifies how the protein-DNA interaction energy is distributed between the association and dissociation transitions. We also assume that the binding to the target is given by  $k^T_{on}=k_{on}$. To test the effect of the sequence symmetry and heterogeneity we consider the protein search on two different types of the DNA molecules: see Figs. 1b and 1c. One of them consists of two homogeneous segments of only $A$ and only $B$ subunits separated by the target (Fig. 1b). Another one is the biopolymer with alternating $A$ and $B$ sites, as presented in Fig. 1c. The block copolymer (Fig. 1b) has a more homogeneous sequence, while the alternating polymers (Fig. 1c) are more heterogeneous. It is important to note that in both cases the overall interaction between the protein and DNA is the same (the overall chemical composition in both cases is identical), and thus our analysis probes {\it only} the effect of the heterogeneity.  This is different from previous computational studies \cite{brackley12}.

\begin{figure} \label{fig:general}
\vspace*{-0.2cm}
\hspace*{-0.1cm}
\centering
\includegraphics[clip,width=0.35\textwidth]{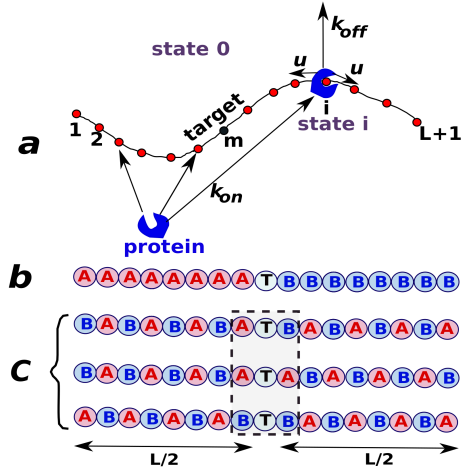}
\caption{ a) A general scheme of the protein search. The DNA chain consists of $L$ nonspecific binding sites and one specific site that is a target for the search. A protein, coming from the solution, can bind to any site on DNA with the association rate per one segment given by $k^{(i)}_{on}$ with $i=A$ or $B$. When attached, the protein can diffuse along the DNA with the rate $u_{i}$  ($i=A$ or $B$), and it can dissociate into the solution with the rate $k^{(i)}_{off}$  ($i=A$ or $B$). The search is finished when the protein binds to the target site at the position $m=L/2+1$. b) A fully symmetric $AB$ block copolymer DNA sequence. c) Pseudo-random alternating sequences with different compositions near the target.}
\end{figure}

To describe the target search dynamics, let us introduce a function $F_n(t)$, which is defined as a first-passage probability to reach the target, if at $t=0$ the protein was at the site $n$ ($n=1, 2, \dots L+1$ corresponds to the starting DNA and $n=0$ is for the bulk solution).  The temporal evolution of this quantity can be described by the  backward master equations \cite{veksler13},
\begin{equation}\label{bm_eq_n}
 \frac{dF_n(t)}{dt}=u_n[F_{n-1}+F_{n+1}]+k^{(n)}_{off}F_0(t) -(2u_n+k^{(n)}_{off})F_{n}(t),  							
\end{equation}
for $1 \le n \le L+1$, while in the solution we have
\begin{equation}
\label{bm_eq_0}
 \frac{dF_0(t)}{dt}=\sum\limits_{n=1}^{L+1}k^{(n)}_{on}F_{n}(t)-F_0(t)\sum\limits_{n=1}^{L+1}k^{(n)}_{on}.
\end{equation}
It is convenient to analyze these equations in the Laplace space using a transformation $\widetilde{F}_n(s)=\int\limits_{0}^{\infty}F_n(t)\,e^{-st}\mathrm{d}t$. Then all probabilities can be found explicitly, which leads to the full dynamic description of the search process. The details of the calculations are presented in the Supplementary Material. More specifically, the mean first-passage time to reach the target starting from the solution is given by  $T_0 \equiv - \frac{\partial \widetilde{F}_0(s)}{\partial s}|_{s=0}$, and other dynamic properties can be also written explicitly. This framework allows us to compare the search dynamics on DNA with different sequences.

In the case of more homogeneous block copolymer sequence (see Fig. 1b), the mean search times are equal to
\begin{equation}
\label{t0_block}
 T_0 = \frac{k_{off}+k_{on}\left[(L/2-P^A)+e^{\varepsilon}(L/2-P^B)\right]}{k_{on}k_{off}(1+P^A+e^{\theta\varepsilon}P^B)},
\end{equation}
where 
\begin{equation}
\label{t0_block_p}
P^{(i)} = \frac{x_i^{1-L/2}-x_i^{1+L/2}}{(1-x_i)(x_i^{1+L/2}+x_i^{-L/2})},
\end{equation} 
\begin{equation}
\label{t0_block_x}
x_i=\frac{2u_{i}+k_{off}^{(i)}-\sqrt{(2u_{i}+k_{off}^{(i)})^2-4u_{i}^2}}{2u_{i}},
\end{equation}
for $i=A$ and $B$. The results are presented in Fig. 2. Again, three dynamic search phases are clearly observed. Increasing the strength of interactions with $B$ subunits make the search in the random-walk regime much slower. This is because the protein gets effectively trapped on $B$ sites for $\lambda > L$.

\begin{figure}\label{fig:phase_diagram_block}
\includegraphics[scale=0.3]{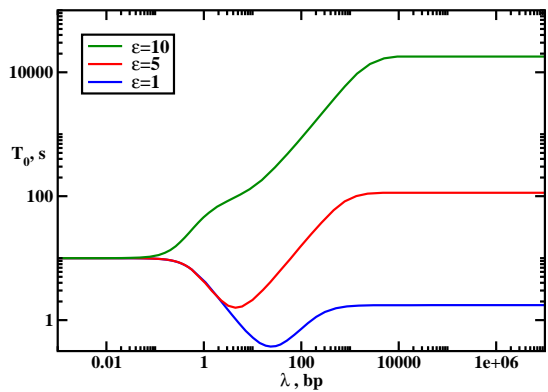}
\caption{Average times to find the target for block copolymer DNA sequence  as a function of the scanning length $\lambda=\sqrt{u/k_{off}}$. The transition rates are: $u=10^5$ s$^{-1}$ and $k_{on}=0.1$ s$^{-1}$. The DNA length is $L=1000$, and we vary the energy difference $\varepsilon$ (in units of $k_{B}T$) for the interaction between the protein and $A$ and $B$ subunits on DNA.}
\end{figure}

Similar expressions for the mean first-passage times can be found for $AB$ alternating  DNA chains, as shown in the Supplementary Material. Here we use the pseudo-random alternating sequences, mimicking the real random situations, because  the analytical results can be obtained for them. But we tested this approximation in computer Monte Carlo simulations by generating random sequences, and one can see from Fig. 3 that this assumption is fully justified. Another interesting observation from Fig. 3 is that the chemical composition near the target might also affect the search dynamics. This can be found only for the intermediate sliding regime ($1 < \lambda < L$) because in this case the probability fluxes to the target site from the solution and from the DNA are comparable. Modifying the composition of the sites near the target can change the amount of the flux coming from the DNA chain.  The flux is larger for $BTB$ sequences (2 $B$ subunits around the target), leading to the smaller search times. This is because the protein molecule attracts stronger to $B$ sites and it has a higher probability to be found here and eventually to go the target. At the same time the flux is smaller for $ATA$ sequences (2 $A$ subunits around the target) with weaker interactions to $A$ sites, which yields slower search dynamics. For $ATB$ sequences, as expected, the intermediate dynamics is observed.

\begin{figure}\label{mc}
\includegraphics[scale=0.3]{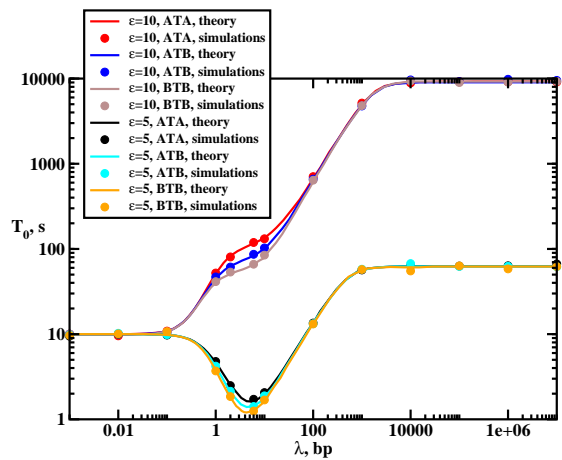}
\caption{Comparison of the search times for alternating sequences with random sequences generated in Monte Carlo computer simulations. The transition rates are: $u=10^5$ s$^{-1}$ and $k_{on}=0.1$ s$^{-1}$. The DNA length is $L=1000$, the loading parameter is $\theta=0.5$, and two different interaction strengths, $\varepsilon=10$ and $\varepsilon=5$, are probed.}
\end{figure}
 
Now we can quantify the effect of sequence heterogeneity in the protein search for the specific binding sites on DNA. The results in Fig. 4 present a ratio of the search times for block copolymer sequences, which are less heterogeneous, and for various alternating sequences, which are more heterogeneous, as a function of the sliding length on DNA. One can see that the effect of the sequence heterogeneity depends on the  nature of the dynamic search phase.  In the jumping regime ($\lambda < 1$), the symmetry of the sequence does not play any role.  This is because in this case the process is taking place only via associations and dissociations (3D search), and the structure of the DNA chain is not important. The situation is different for the intermediate sliding regime (3D+1D search, $ 1 < \lambda < L$) where in most cases the search on alternating sequences is faster. This can be explained by noting that the search time in this dynamic phase is proportional to $L/\lambda$ \cite{veksler13}, which gives the average number of cycles before the protein can find the target. In the block copolymer sequence the protein mostly comes to the target from the $B$ segment because of stronger interactions. In the alternating sequences the protein can reach the target from both sides. It can be shown analytically (see the Supplementary Material) that the scanning length on the alternating segment is larger than the scanning length for the $B$ segment, i.e., $\lambda_{AB} > \lambda _{B}$. Then the search time is obviously faster for the alternating sequence because $ L/\lambda_{AB} < L/\lambda_{B}$. The only deviation from this picture is found in $ATA$ sequences where for small range of parameters the search is slower than in the block copolymer sequence. The effect of the chemical composition near the target, as discussed above, is responsible for this. In the random-walk  regime (1D search, $\lambda > L$), the effect of the sequence heterogeneity is even stronger: the protein molecule finds the specific binding site  up to 2 times faster  for more heterogeneous DNA chains. To understand this behavior, we note that in this case the mean first-passage time to reach the target is a sum of residence times on the DNA sites. Because the target is in the middle of the chain, the mean time to reach the target from the block copolymer sequence will be $T_{0} \simeq (L/4) \tau_{B}$, where $\tau_{B}$ is the residence time at the site $B$. The average starting position of the protein is $L/4$ sites away from the target.  For the alternating sequences, the average distance to the target is the same, but the chemical composition of intermediate sites is different, yielding, $T_{0} \simeq (L/8) \tau_{B} +(L/8) \tau_{A}$. Obviously, the protein spends much less time on $A$ subunits, and this leads to faster search for alternating DNA sequence. For $\tau_{A} \ll \tau_{B}$ this also explains the factor of 2 in the search speed. In this case, the $B$ subunits can be viewed as traps. Thus, in dynamic phases where the structure of DNA is important the sequence heterogeneity almost always accelerates the protein search for targets.

\begin{figure}\label{fig:ratio_alter_block}
\includegraphics[scale=0.3]{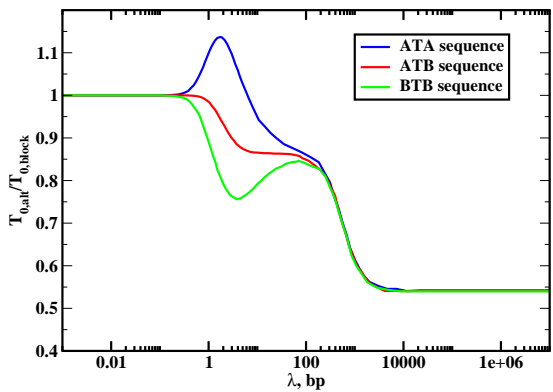}
\caption{ The ratio of the search times  for the alternating DNA sequences and for the block copolymer DNA sequences as a function of the scanning length $\lambda=\sqrt{u/k_{off}}$. Three different chemical compositions near the target are distinguished, namely, $ATA$, $ATB$, $BTB$. The transition rates are: $u=10^5$s$^{-1}$ and $k_{on}=0.1$ s$^{-1}$. The DNA length is $L=1000$, the loading parameter is $\theta=0.5$, and the energy difference of interactions for the protein with $A$ and $B$ sites is $\varepsilon=5$. }
\end{figure}

In conclusion, we presented a theoretical analysis of DNA sequence symmetry and heterogeneity in the protein search process. Using analytical solutions of the discrete-state stochastic approach that accounts for most important physical-chemical processes in the system, we obtained a full description of the search dynamics. It is found that the sequence heterogeneity is a crucial factor in the facilitated diffusion. Unlike the previous theoretical and computational models, our approach predicts that the sequence heterogeneity mostly accelerates the search. The mechanisms of this phenomenon depend on the nature of the search regime. It is either the smaller number of search cycles or the smaller number of trapping sites on the path to the target. We also found that in the dynamic phase where the specific binding site can be reached from  the solution and from the DNA chain, the chemical composition near the target might influence the search dynamics. The search is faster if the target is surrounded by the subunits which interact stronger with the protein, providing it more opportunities to reach the target. Our theoretical results not only clarify the fundamental physics of the protein search dynamics, but they also suggest that the biological processes can be effectively regulated by modifying  the sequence symmetry and heterogeneity in DNA, as well as the chemical composition near the targets. Experiments to test these predictions should provide a better understanding of the microscopic mechanisms of complex biological processes.

The work was supported by the Welch Foundation (Grant C-1559), by the NSF (Grant CHE-1360979), and by the Center for Theoretical Biological Physics sponsored by the NSF (Grant PHY-1427654).

\end{document}